\documentclass[onecolumn,preprint,amsmath,amssymb]{revtex4-1}
\usepackage{graphicx}
\usepackage{dcolumn}
\usepackage{mathrsfs}
\usepackage{bm}
\usepackage{lineno}

\begin{document}

\title{Optimum quantum resource distribution for phase measurement and quantum information tapping in a dual-beam SU(1,1) interferometer}

\author{Yuhong Liu$^{1}$}
\author{Nan Huo$^{1}$}
\author{Jiamin Li$^{1}$}
\author{Liang Cui$^{1}$}
\author{Xiaoying Li$^{1,*}$}
\author{Z. Y. Ou$^{1, 2,\dag}$}
\affiliation{%
$^{1}$College of Precision Instrument and Opto-Electronics Engineering, Key Laboratory of
Opto-Electronics Information Technology, Ministry of Education, Tianjin University,
Tianjin 300072, P. R. China\\
$^{2}$Department of Physics, Indiana University-Purdue University Indianapolis, Indianapolis, IN 46202, USA
}%

\date{\today}




\begin{abstract}
Quantum entanglement is a resource in quantum metrology that can be distributed to two orthogonal physical quantities for the enhancement of their joint measurement sensitivity, as demonstrated in quantum dense metrology. On the other hand, we can also devote all the quantum resource to phase measurement only for optimum measurement sensitivity. Here, we experimentally implement a dual-beam scheme in an SU(1,1) interferometer for the optimum phase measurement sensitivity. We demonstrate a 3.9-dB improvement in signal-to-noise ratio over the optimum classical method and this is 3-dB better than the traditional single-beam scheme. Furthermore, such a scheme also realizes a quantum optical tap of quantum entangled fields and has the full advantages of an SU(1,1) interferometer for practical applications in quantum metrology and quantum information.
\end{abstract}


\maketitle

Phase measurement sensitivity has been a topic of constant interest ever since optical interferometry technique was invented more than one hundred years ago \cite{mich}. The employment of quantum states of light in interferometry has now pushed the measurement sensitivity to a new limit, beyond what is allowed with classical coherent sources of light \cite{cav81,gio}. Squeezed states, because of the property of quantum noise reduction, are usually applied to a traditional interferometer for sensitivity enhancement in phase measurement \cite{cav81,xiao,gran}. Quantum entanglement, as a quantum resource, can also be applied to enhance phase measurement sensitivity by quantum noise cancelation via quantum correlation \cite{bra00,ligoepr}.

Recently, a new type of quantum interferometer known as SU(1,1) interferometer was demonstrated to exhibit sensitivity enhancement in phase measurement \cite{yur,pl10,jing11,ou12,lett12,hud14,lett17,che17} and in the meantime possesses detection loss tolerance property \cite{ou12,hud14,che17}, which is a huge advantage over the squeezed state schemes. Although the hardware of the new interferometer changes from beam splitters to parametric amplifiers, the underlining physics is still quantum noise reduction by noise cancelation through quantum entanglement \cite{kong13,lett17,and17, JML}, similar to Ref.\cite{bra00}.
However, it was shown \cite{JML} that these quantum entanglement-based schemes can only gives rise to half the sensitivity enhancement in phase measurement as compared to the squeezed state schemes with the same gain parameters in the parametric processes for their generation. The study discovered \cite{JML} that these schemes are also able to increase the sensitivity of the amplitude measurement concurrently with the phase measurement under the name of ``quantum dense metrology" \cite{li02,snb13,liu_OE_2018}. Therefore, the quantum resource of entanglement is split between phase and amplitude measurement.

To increase the phase measurement sensitivity to optimum, on par with the squeezed state schemes, we need to devote all the quantum resource to phase measurement only.
In this letter, we experimentally implement a variation of the SU(1,1) interferometer which employs both the signal and the idler beams to probe a common phase shift. We find this dual beam sensing scheme can double the sensitivity of the original single-beam sensing scheme, making full use of the quantum resource of entanglement for phase measurement. Furthermore, such a scheme also achieves for the first time quantum tapping of information encoded in quantum entangled fields.

\begin{figure}[htbp]
\centering
\includegraphics[width=10cm]{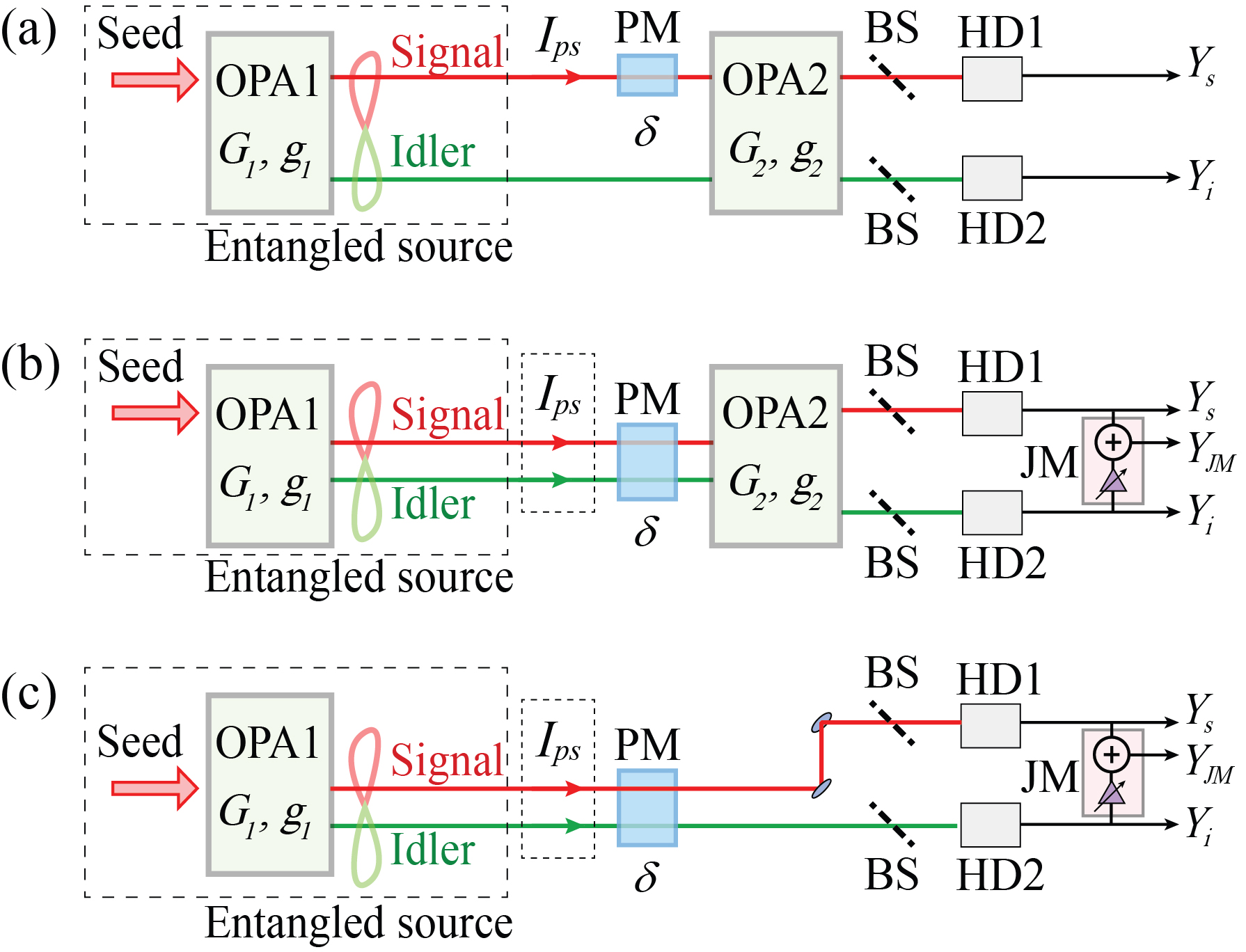}
\caption{
Phase measurement schemes with entangled source from an optical parametric amplifier (OPA1).
(a) Single-beam sensing SU(1,1) interferometer;
(b) Dual-beam sensing SU(1,1) interferometer;
(c) Direct joint measurement scheme.
PM, phase modulator;
HD, homodyne detection;
\(I_{ps}\), phase sensing field intensity.
BS, beam splitter with transmission efficiency of $1-\eta$ for modeling non-ideal detection efficiency.
}
\label{fig2:compare}
\end{figure}

Quantum enhanced phase measurement was recently achieved with a new type of quantum interferometer, that is, the SU(1,1) interferometer (SUI), which utilizes parametric amplifiers (OPA1,OPA2), instead of traditional beam splitters, for wave splitting and superposition, as shown in Fig. \ref{fig2:compare}(a).
Variations of SUI include the scheme with OPA2 replaced by a beam splitter \cite{kong13} and a truncated scheme where homodyne measurements are performed directly on the modulated beam, and the resulting photo-currents are added or subtracted for quantum noise cancellation \cite{lett17,and17,gup18}.
It was shown \cite{JML} that all these schemes have an optimized phase measurement sensitivity characterized by the signal-to-noise ratio (SNR) in the measurement of Y-quadrature at the outputs:
\begin{eqnarray}
SNR_{SUI} = 2I_{ps}\delta^2(G_1+g_1)^2,
\label{eq:SNR-SUI}
\end{eqnarray}
where $\delta$ is the phase modulation on the probe beam, $I_{ps}$ is the photon number of the phase sensing field, and $G_1,g_1$ are the amplitude gains of the first parametric amplifier, which creates entanglement between the two arms of the interferometer.
Eq. (\ref{eq:SNR-SUI}) is obtained when the gain amplitude of second optical parametric amplifier (OPA2) $G_2$ is much higher than that of first amplifier OPA1, i.e., $G_2\gg G_1$. This has an improvement factor of $(G_1+g_1)^2/2$ over the optimum classical sensitivity of
\begin{eqnarray}
SNR_{HD} = 4I_{ps}\delta^2
\label{eq:SNR-HD}
\end{eqnarray}
by a classical probing field, which is achieved by homodyne detection (HD) \cite{gup18,JML}. It should be mentioned that previously derived classical SNR \cite{ou12,ligo09,Ligo77} is not optimized and only is half the SNR in Eq.(\ref{eq:SNR-HD}).
The reason for this difference was discussed by Gupta {\it et al.} \cite{gup18}. In the following, we will compare all our results to the optimum classical SNR in Eq.(\ref{eq:SNR-HD}) (SQL2 in Ref.\cite{gup18}).

However, it is well-known that with a coherent squeezed state as the probe, quantum noise can be reduced and the SNR can be improved to
\begin{eqnarray}
SNR_{SQ} = 4|\alpha|^2\delta^2/S = 4I_{ps}\delta^2(G_1+g_1)^2
\label{eq:SNR-SQ}
\end{eqnarray}
where $S\equiv 1/(G_1+g_1)^2$ is the squeezed quantum noise and $G_1, g_1$ are similar to those in Eq.(\ref{eq:SNR-SUI}) but are the amplitude gains for the degenerate parametric amplifier generating the squeezed state. $SNR_{SQ}$ is enhanced
by a factor of $1/S \equiv (G_1+g_1)^2$ as compared to the optimum classical SNR in Eq.(\ref{eq:SNR-HD}). This is a factor of two larger than that in Eq.(\ref{eq:SNR-SUI}) for the SUI scheme. Notice that squeezing $S$ is extremely sensitive to transmission and detection losses, which degrade the enhancement factor to $1/S'=1/[S+\eta/(1-\eta)]$ with $\eta$ as the overall loss, modeled as a beam splitter (BS) with transmission $1-\eta$.

The reason for the difference between the two results in Eqs.(\ref{eq:SNR-SUI}) and (\ref{eq:SNR-SQ}) is quantum resource sharing. It was shown recently \cite{JML,liu_OE_2018} that the scheme in Fig. \ref{fig2:compare}(a) can also be used to measure amplitude modulation on the probe beam with the measurement of X-quadrature (by HD2) at the other output port of the second amplifier of the SU(1,1) interferometer with the same SNR given in Eq. (\ref{eq:SNR-SUI}). Since the phase measurement and the amplitude measurement are performed at different ports of the OPA2, they can be done simultaneously, sharing the same resource of quantum entanglement \cite{JML}. The quantum resource is thus split between the phase and amplitude measurement, reducing the quantum enhancement effect by half for each measurement.

Realizing this difference, we now construct a variation of SUI to devote all the quantum resource to phase measurement.
The new scheme involves both correlated signal and idler fields for probing the phase change, as shown in Fig. \ref{fig2:compare}(b).
A quick comparison of the new scheme with a traditional interferometer such as a Mach-Zehnder interferometer may lead to the concern that the phase change signal would be canceled in phase difference, a phenomenon known as ``common mode rejection" in a traditional interferometer.
Fortunately, the working principle of SUI is totally different from that of a traditional interferometer: the interference output depends on the sum of the phases of the arms rather than the difference \cite{jing11,Chen15}.
Therefore, the size of phase signal $\delta$ will be doubled as compared to the single-beam SUI scheme in Fig. \ref{fig2:compare}(a), but $I_{ps}$ will also double because of the dual-beam probing (under the condition of $G_1\gg 1, G_1\approx g_1$). Overall, there is an increase of the SNR by a factor of 2, recovering the SNR in Eq. (\ref{eq:SNR-SQ}).
Indeed, a straightforward calculation for the dual-beam SUI scheme in Fig. \ref{fig2:compare}(b) finds that when $G_2\rightarrow \infty$,
the optimum SNR of phase signal measured by homodyne detection (HD) of Y-quadrature at each output of SUI is given by \cite{JML}:
\begin{eqnarray}
SNR_{HD1}^{(PM)} = SNR_{HD2}^{(PM)} = \frac{2(G_1+g_1)^4I_{ps}\delta^2}{G_1^2+g_1^2},
\label{eq:SNR-Amp-op}
\end{eqnarray}
where $I_{ps}\equiv (G_1^2+g_1^2)|\alpha|^2$ ($|\alpha|^2\gg 1$).
When $g_1\gg 1$, the results in Eq. (\ref{eq:SNR-Amp-op}) are
exactly the same as Eq. (\ref{eq:SNR-SQ}), which is twice of that in Eq. (\ref{eq:SNR-SUI}) for single beam sensing scheme. On the other hand, the ``common mode rejection effect" does apply to amplitude modulation, leading to canceled signal size at either the outputs of the second amplifier and reduced SNR \cite{JML}:
\begin{eqnarray}
SNR_{HD2}^{(AM)} = 2I_{ps}\epsilon^2 /(G^2_1+g_1^2)\sim 0~{\rm for}~g_1\gg1,
\label{eq:SNR-Amp-op3}
\end{eqnarray}
where $\epsilon$ is the amplitude modulation on dual-beam. So, this scheme is not good for amplitude measurement.
Notice that with $\delta=\epsilon$, we have
\begin{eqnarray}
SNR_{HD1}^{(PM)} + SNR_{HD2}^{(AM)}&=& 4(G_1+g_1)^2I_{ps}\delta^2 \cr
&=& SNR_{SQ}(PM).
\label{eq:SNR-Amp-sum}
\end{eqnarray}
Note the right hand side of Eq.(\ref{eq:SNR-Amp-sum}) is the same as Eq.(\ref{eq:SNR-SQ}).
This demonstrates the quantum resource sharing between the measurement of
conjugate variables (phase and amplitude) for arbitrary $g_1$.
From this analysis, we find that the quantum resource can be all used for phase measurement when $g_1\gg1$ and this scheme recovers the enhancement factor lost due to quantum resource sharing.

It should be noted that
in addition to the homodyne detection measurement on the quadrature amplitude at each ouputs of OPA2, $Y_{HD1}$ or $Y_{HD2}$,
joint measurement $\hat Y_{HD1}+\hat Y_{HD2}$ can be performed by adding up the photo-currents out of HD1 and HD2.
Although optimum performance of SUI can be achieved for the measurement of $\hat Y_{HD1}$ or $\hat Y_{HD2}$ when $G_2\rightarrow\infty, G_1\gg 1$, it is shown that the joint measurement can give rise to the optimum quantum enhancement in Eq. (\ref{eq:SNR-Amp-op}) by properly adjusting the electronic gain parameter in the joint measurement circuit even at finite gains of $G_2$ \cite{JML18}.

For the completeness of discussion and later discussion on quantum information tapping, we consider a direct detection dual-beam scheme shown in Fig. \ref{fig2:compare}(c) where both the entangled signal and idler fields from OPA1 are used to probe the phase shift.
This scheme is similar to the truncated SU(1,1) interferometers in Refs.\cite{lett17,and17,gup18}, where OPA2 is replaced by a current mixer to obtain the joint quantity $ i_{JM} = i_s+k_i i_i$ with $i_s$ and $i_i$ as the photo-currents directly from HD1($\hat Y_s$) and HD2($\hat Y_i$), respectively.
However, we utilize the dual-beam sensing here instead of the single-beam sensing in Refs.\cite{lett17,and17,gup18}. It is straightforward to show that the SNR in the measurement of the joint quantity $ i_{JM}$ is the same as that given in Eq. (\ref{eq:SNR-SQ}) when the gain parameter $k_i$ takes the optimized value of $k_i=1$\cite{JML,JML18}.
However, just like the squeezed state scheme given in Eq.(\ref{eq:SNR-SQ}), this scheme is sensitive to detection losses and the enhancement degradation has exactly the same dependence on losses as the squeezed state scheme discussed earlier. For the SU(1,1) interferometers in Fig. \ref{fig2:compare}(a,b), however, it is another story \cite{JML}.
Even with some detection losses (denoted as $\eta$ before detection), the improvement factor only changes to $1/S'' = 1/[S+\eta/2G^2_2(1-\eta)] $, which is approximately $1/S$ when $G_2 \gg G_1$ for the scheme in Fig. \ref{fig2:compare}(b).
So, the scheme of SUI using OPA2 to coherently combine the entangled signal and idler fields is tolerant to detection losses. This is because the output noise of OPA2 is much larger than vacuum noise so that the vacuum noise coupled in through loss channel is negligible.

Another interesting application of the dual-beam sensing scheme in Fig. \ref{fig2:compare}(b) is the realization of a quantum information tap \cite{sha80,lev93b}. In this case, OPA2 is regarded as the information splitter for the input, which is the entangled fields from OPA1 with a phase signal encoded by PM. The input SNR corresponds to the direct joint measurement result in Fig. \ref{fig2:compare}(c) and is the same as that given in Eq.(\ref{eq:SNR-SQ}): $SNR_{in} = SNR_{SQ}$ for the lossless case. With output SNRs at the two output of OPA2 given in Eq. (\ref{eq:SNR-Amp-op}), we have the optimum transfer coefficients as
\begin{eqnarray}
T_{HD1,HD2} &\equiv & \frac{SNR_{HD1,HD2}^{(PM)}}{SNR_{in}} = \frac{(G_1+g_1)^2}{ 2(G_1^2+g_1^2)} \cr
&\approx & 1 ~~{\rm for}~~ g_1\gg 1.
\label{eq:SNR-T}
\end{eqnarray}
So, $T_{HD1}+T_{HD2}= (G_1+g_1)^2/(G_1^2+g_1^2) >1$, satisfying the condition for quantum optical tapping \cite{sha80}: the signal encoded in the entangled fields is split into two by the amplifier without adding noise in the ideal case of $g_1\gg 1$.

\begin{figure*}[htbp]
\centering
\includegraphics[width=17cm]{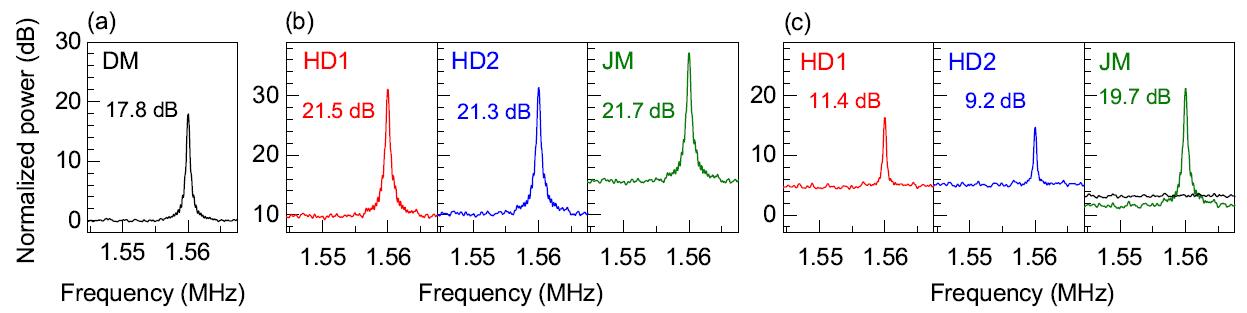}
\caption{The spectrums from HD1, HD2 and joint measurement (JM) for the measurement of a phase modulation signal at 1.56 MHz. The noise levels are all normalized to the shot noise level of HD1. (a) Direct measurement by HD1 with a coherent probe beam when gains of OPA1 and OPA2 are one; (b) measurement from SUI when the power gains of OPA1 and OPA2 are 2.5 and 12; (c) measurement with entangled probe beams (truncated SUI scheme) when the power gains of OPA1 and OPA2 are 2.5 and 1. The black line in JM of (c) is the shot noise level $SNR_{si}$ of the joint quantity $\Delta i = i_s+k_ii_i$ ($k_i=1$), which is exactly 3 dB above the shot noise level at 0 dB for individual HD1.}
\label{fig4:spectrum}
\end{figure*}

We implement experimentally the dual-beam schemes in Fig. \ref{fig2:compare}(b,c) with fiber-based parametric amplifier. The detail of the experimental setup is given in Supplementary Materials.
A typical data set is presented in Fig. \ref{fig4:spectrum}. Fig. \ref{fig4:spectrum}(a) shows the result of HD1 for the optimum classical phase measurement scheme when the gains of OPA1 and OPA2 are set to one and they act simply as transmission media. The peak at 1.56 MHz corresponds to the measured power of phase modulation signals. One sees that the SNR is $17.8\pm 0.2$ dB, which is the benchmark SNR for the classical phase measurement that we will compare to. Note that the noise levels in all the plots in Fig. \ref{fig4:spectrum} are normalized to the shot noise level of HD1 for the sake of easy comparison.

Fig. \ref{fig4:spectrum}(b) shows the result of dual-beam sensing SUI. In this measurement, the power gains of OPA1 and OPA2 are 2.5 and 12, respectively. From the measurement of HD1 and HD2, which are the individual homodyne detections at signal and idler outputs, respectively, one sees that the SNRs of phase signal at 1.56 MHz reads $21.5\pm 0.2$ dB and $21.3\pm 0.2$ dB. These are $3.7\pm 0.3$ dB and $3.5 \pm 0.3$ dB improvement over the classical measurement result in Fig. \ref{fig4:spectrum}(a). From the joint measurement (JM) of HD1 and HD2 with an optimized electronic gain $k_i=1$, we find the SNR of measured phase signal is $21.7\pm 0.2$ dB, which corresponds to an improvement of $3.9 \pm 0.3$ dB over the classical result. If the classical limit of the phase measurement were defined the same as that of the single-beam sensing SUI reported in Ref.\cite{Guo2016,liu_OE_2018}, the improvement obtained for the JM case in Fig. \ref{fig4:spectrum}(b) would be 6.9 dB. This result therefore demonstrates the advantage of dual-beam sensing SUI over the single-beam SUI scheme. Notice that the joint measurement has an SNR slightly higher than the SNRs from individual measurement of HD1 or HD2. This is because the gain of OPA2 are finite in our experiment. So the SNRs at each output of SUI are not optimized as discussed in Eq.(\ref{eq:SNR-Amp-op}), whereas the joint measurement always gives the optimized value (only determined by the entanglement degree) irrespective of the gain of OPA2 .

To compare the performance of dual-beam sensing SUI with the traditional squeezed state scheme by direct detection, we implement the dual-beam phase measurement scheme in Fig.\ref{fig2:compare}(c) with entangled source. In this experiment, the gain of OPA2 is set to one but the gain of OPA1 and the incident seed are the same as the case in Fig. \ref{fig4:spectrum}(b). The results obtained by HD1 and HD2 and joint measurement (JM) are presented in Fig. \ref{fig4:spectrum}(c). Although SNRs extracted from HD1 and HD2 individually are much smaller than the classical measurement result in Fig. \ref{fig4:spectrum}(a) due to thermal nature of the individual signal and idler fields, the joint measurement gives an SNR of $19.7\pm 0.2$ dB, which is $1.9 \pm 0.3$ dB better than the SNR in Fig. \ref{fig4:spectrum}(a). Indeed, this scheme corresponds to the case of truncated SU(1,1) interferometer \cite{lett17,gup18}, where the role of OPA2 is replaced by a current mixer of HD1 and HD2 for superposition of the signal and idler fields. The joint measurement of HD1 and HD2 with $k_i=1$ has the modulated signal coherently added (nearly 6 dB increase) but noise reduced below the joint shot noise level $SNL_{si}$ (see the black line in Fig.\ref{fig4:spectrum}(c)) due to noise anti-correlation between $Y_s$ and $Y_i$ of the entangled signal and idler fields \cite{Guo-OL16}. It is interesting to note that had we used the single-beam scheme in Ref.\cite{lett17,gup18}, the observed SNR would be about 3 dB (factor of 2) smaller and we would have worse SNR (about -1 dB) than the optimum classical result shown in Fig. \ref{fig4:spectrum}(a).
Moreover, the 1.9 dB improvement shown in Fig. \ref{fig4:spectrum}(c) over the optimum classical result is lower than the 3.9 dB improvement shown in Fig. \ref{fig4:spectrum}(b). This is because direct joint measurement of quantum entanglement is prone to propagation and detection losses (about 25\%) in our system. The extra vacuum noise from losses will reduce the effect of quantum correlation and quantum noise reduction. The SUI scheme, on the other hand, is insensitive to these losses because each output of OPA2 has both the phase signal and noise amplified with noise level well above loss-induced vacuum noise for $G_2 \gg G_1$. Losses will then decrease the signal and noise level at the same ratio~\cite{JML18}, leading to no obvious change in SNR. Hence, this comparison shows the loss-tolerant property of SUI.

On the other hand, Fig. \ref{fig4:spectrum}(b) also demonstrates the realization of a quantum optical tap by OPA2. The input fields to OPA2 are the two entangled fields generated from OPA1 (see Fig.\ref{fig2:compare}(c)), which serves as the quantum signal to be split. The direct joint measurement of phase signal carried by entangled signal and idler beams gives an SNR of $19.7\pm 0.2$ dB, as shown in Fig. \ref{fig4:spectrum}(c), which is $1.9 \pm 0.3$ dB better than the SNR obtained by classical phase measurement in Fig. \ref{fig4:spectrum}(a) due to noise reduction originated from the quantum correlation of two entangled fields. However, as we mentioned in theory part, the direct joint measurement scheme is sensitive to losses in transmission and detection. After correction of these losses on both the noise level and the modulation signal size, the SNR of phase signal right at the output of PM is $22.9 \pm 0.2$ dB. So we take SNR of input quantum signal as $SNR_{in} = 22.9 \pm 0.2$ dB.
The split fields are the signal and idler outputs of OPA2, which are respectively measured by HD1, HD2. From the measurement of HD1 and HD2 shown in Fig. \ref{fig4:spectrum}(b), we find the SNRs of two outputs are $SNR_s=21.5 \pm 0.2$ dB and $SNR_i=21.3 \pm 0.2$ dB, respectively. These results lead to transfer coefficients of $T_s=SNR_s/SNR_{in} = 0.72\pm 0.06$ and $T_i=SNR_i/SNR_{in} = 0.69\pm 0.06$, with $T_s+T_i=1.41\pm 0.09$, which is larger than the classical limit of 1. The reason that the transfer coefficients are different from the ideal ones is two folds. One is the finite gain of OPA2, the other is the transmission loss and mode mismatching loss occurred when the entangled signal and idler beams are coupled into OPA2.

In summary, we construct a dual-beam sensing SUI and demonstrate its advantages in both phase measurement sensitivity and loss tolerance. The measurement results show that SNR is $3.9 \pm 0.3$ dB higher than that obtained by the optimum classical method in phase measurement. Using the dual-beam sensing SUI, we also realize for the first time the information splitting of phase signal encoded on entangled fields with high transfer coefficients of $0.72\pm 0.06$ and $0.69\pm 0.06$, respectively, satisfying the condition for quantum optical tapping.

Compared to previous methods, dual-beam sensing SUI not only makes full use of the quantum resource for phase measurement, but also is insensitive to propagation and detection losses and thus lifts the barrier for the quantum enhanced metrology and quantum communication in practical applications.
The loss tolerance property shows that dual-beam sensing SUI has great potentials in those situations when quantum efficiency of detection system limits the implementation of quantum enhanced measurement, such as those working at wavelength that lacks efficient photo-detectors (for example, wavelength longer than 2 $\mu$m or ultra violet region).

Although the dual-beam sensing scheme has twice the SNR as the single-beam sensing scheme, its implementation requires the two correlated beams to be nearly the same so as to probe the same phase change. Very often SUI is realized with different types of waves as in the atom-light hybrid interferometer \cite{Chen15} where the phases involved belong to light and atom separately. In this case the dual beam scheme wouldn't work.

\acknowledgments{
This work is supported by
the National Key Research and Development Program of China (2016YFA0301403),
the 973 program of China (2014CB340103),
the National Natural Science Foundation of China (91736105, 11527808),
and the 111 project (B07014), and by US National Science Foundation (No.1806425).}

\vskip 0.1in

\noindent $*$ xiaoyingli@tju.edu.cn

\noindent $\dag$ zheyuou@tju.edu.cn

\newpage

\newpage

\section{Supplementary Materials: Experimental setup}

\begin{figure*}[htbp]
\centering
\includegraphics[width=15cm]{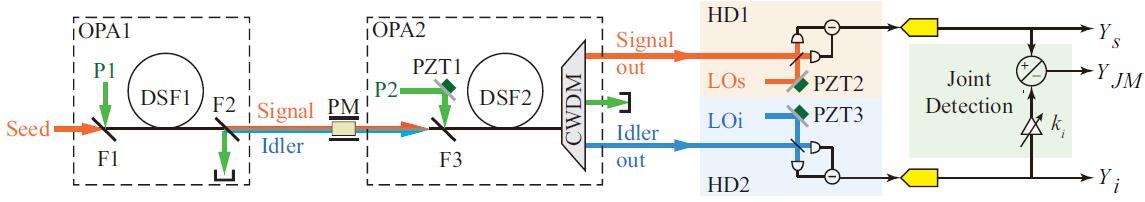}
\caption{Experiment setup.
DSF1-2, dispersion-shifted fiber optical parametric amplifiers;
F1-3, band-reflection filter centering at 1550nm;
CWDM, coarse wavelength division multiplexer;
HD1-2, homodyne detectors;
LOs(LOi), local oscillator for signal (idler) field;
PZT, piezo-electric transducer;
\(k_i\), electronic variable gain
 }
\label{fig2:setup}
\end{figure*}

The experimental setup for measuring the weak phase modulation by using the dual-beam sensing SUI is shown in Fig. \ref{fig2:setup}. There are two fiber-based OPAs in the scheme. The nonlinear media for OPA1 and OPA2 are two pieces of identical dispersion shifted fibers (DSF1 and DSF2).
The length and zero dispersion wavelength of each DSF are about 150 m and 1548.5 nm, respectively.
The pumps for each OPA, P1/P2, is a mode locked pulse train.
Each pulsed pump with full width at half maximum (FWHM) of 0.4 nm is centered at 1549 nm to ensure the phase matching of four wave mixing (FWM) parametric process is satisfied in DSF~\cite{Darwin-OL99}.
OPA1 generates the entangled signal and idler beams~\cite{Guo-APL12,Guo-OL16}.
When the strong pump P1 and weak seed injection centering at 1533 nm are combined by a wavelength division multiplexer (WDM) filter (F1) and simultaneously launched into DSF1, we obtain the amplified signal beam and generated idler beam via FWM.
By passing the output of OPA1 through a WDM filter (F2), we isolate the residual power of pump P1 and select out the entangled signal and idler beams, which are centering at 1533 and 1566 nm, respectively, and co-propagate in space. 
We then encode a weak phase signal by propagating both the signal and idler beams through a phase modulator modulated at the frequencies of 1.56 MHz.
The encoded dual-beam is combined with the pump P2 and simultaneously launched into DSF2 for signal amplification and noise suppression.
At the output of OPA2, we exploit a 2-channel coarse wavelength division multiplexer (CWDM) to isolate the pump P2 and to efficiently select out the signal and idler fields.
The isolation degree for the two channels of CWDMs, which possess high transmission efficiencies for the signal and idler fields, respectively, is greater than 40 dB.
At the two outputs of OPA2, the signal and idler fields are respectively measured by the homodyne detection systems HD1 and HD2.
The local oscillator of (LOs/LOi) of each homodyne detection HD1/HD2 is properly locked to measure the quadrature phase $\hat Y_{s/i}$.
For the joint measurement, we then mix the photo-currents of HD1 and HD2 with a mixer, and the output of HD2 is adjusted with an electrical gain $k_i$ to optimize the measured SNR.
The power spectrum of measurement result is analyzed by sending the photon-currents directly out of each individual homodyne detector HD1 and HD2 ($i_s$, $i_i$) and joint measurement ($i_{JM}$) to a data acquisition system (DAQ) for spectral analysis.

To ensure that the phase signal encoded on the dual-beam can be measured at each output of SUI with optimum SNR, the power gain of OPA2 should be much higher than that of OPA1.
During the measurement, the powers of P1 and P2 are 2 mW and 4 mW, respectively.
Under this condition, the power gains of OPA1 and OPA2 are about 2.5 and 12, respectively.
Moreover, to obtain the best noise cancelation effect, OPA2 is operated at the dark fringe point by locking the relative phase between its pump P2 and two inputs.

To clearly demonstrate the quantum enhancement, we need to compare the SNR measured by the dual-beam sensing SUI with that obtained by using classical method.
The classical method, corresponding to a direct homodyne detection, is realized by setting the powers of two pumps (P1 and P2) to zero so that the two DSFs are simply transmission media.
On the other hand, to illustrate the loss tolerance feature, we compare the measurement results of dual-beam sensing SUI with that obtained by direct joint measurement of the entangled fields out of OPA1 (P1 is on but P2 is zero), which is equivalent to the scheme in  Fig. 1(c).
It is worth noting that in each case, the probe beam intensity $I_{ps}$ is adjusted to be the same for fair comparison.
The presented measurement results  are all obtained with $I_{ps}=200$ pW.

In the experiment, the two pumps (P1 and P2) are obtained by carving the output of a femto-second laser with repetition rate and central wavelength of about 36.9 MHz and 1550 nm, respectively.
The preparation of other optical fields, including the seed injection and local oscillators of HDs, and the realization of mode matching between two OPAs are described in Ref. \cite{  liu_OE_2018}.
The technical details for locking the phase of OPA2 and two sets of HDs by loading the feedback signals on PZTs are given in our previous publications (see Refs. \cite{liu_OE_2018} and \cite{Guo2016}).

The transmission efficiency between OPA1 and OPA2 is about 70\%. The total detection efficiency of the signal/idler output is about 78\%/73\%.

%


\end{document}